\documentclass[12pt]{article}

\newcommand{\ket}[1]{|#1\rangle}
\newcommand{\inket}{|\mbox{in}\rangle}
\newcommand{\outket}{|\mbox{out}\rangle}
\newcommand{\bra}[1]{\langle#1|}

\begin{document}

\begin{center}
{\LARGE Losing Your Marbles  in \\ Wavefunction Collapse 
Theories\footnote{Forthcoming in \emph{The British Journal for 
Philosophy of Science}, December 1999.} 
\par} 
\lineskip .75em
{\Large \emph{Rob Clifton and Bradley Monton} \par} \vskip .75em
\end{center}
\begin{center}
\Large ABSTRACT 
\lineskip .75em
\end{center}
Peter Lewis ([1997]) has recently argued that the wavefunction collapse theory of GRW 
(Ghirardi, Rimini, and Weber [1986]) can only solve the problem of 
wavefunction tails at the expense of predicting that arithmetic 
does not apply to ordinary macroscopic objects.  More specifically, 
Lewis 
argues that the GRW theory must violate
\emph{the enumeration principle}: that `if marble 1 is in the box and marble 2 is in the box and so on
through marble $n$, then all $n$ marbles are in the box' ([1997], p. 321).  Ghirardi and
Bassi ([1999]) have replied that it is meaningless to say that 
the enumeration principle is 
violated because the wavefunction Lewis uses to exhibit the 
violation cannot persist, according to the GRW theory, for more than a 
split second ([1999], p. 709).  
On the contrary, we argue that Lewis's argument survives Ghirardi and Bassi's 
criticism unscathed.  We then go on to show that, while the enumeration 
principle \emph{can fail} in the GRW theory, the theory itself guarantees 
that the principle \emph{can never be empirically falsified}, leaving the applicability of 
arithmetical reasoning to both micro- and macroscopic
objects intact.
\vskip .5em
\vskip .5em
\vskip .5em
\noindent 1 \textit{Wavefunction Collapse Theories and the Tails Problem}
\vskip .5em
\noindent 2 \textit{Lewis's Counting Anomaly}
\vskip .5em
\noindent 3 \textit{Can the Counting Anomaly Be Avoided?}
\vskip .5em
\noindent 4 \textit{Is the Counting Anomaly Ever Manifest?}
\vskip .5em
\noindent 5 \textit{Is Suppressing the Manifestation of Anomalies 
Enough?}
\newpage

\section{Wavefunction Collapse Theories and the Tails Problem} 
The standard Schr\"{o}dinger dynamics for a quantum
system prescribes that its state vector
$\ket{\psi(t)}$ always evolves in time deterministically,
and linearly (i.e., that $\ket{\psi(t)}$'s evolution is the sum of the separate evolutions of its
components in
any basis).   The standard `eigenstate-eigenvalue link' \emph{semantics}  for quantum states
dictates that
 an observable $O$ of a quantum
system in state $\ket{\psi(t)}$ possesses a determinate value at time $t$ if and only if
$O\ket{\psi(t)}=o\ket{\psi(t)}$ for some eigenvalue $o$ of $O$ (i.e., if and only if the probability
of finding $o$ in a measurement of $O$ is 1 at time $t$).  Unfortunately, the
standard dynamics and semantics for quantum states together give rise to the
measurement problem; they force the conclusion that a cat can be neither alive nor dead,
and, worse, that a competent observer who looks
upon such a cat will neither believe that the cat is alive nor believe it to be dead.
The standard way out of the measurement problem is
to keep the
standard semantics and temporarily suspend the standard dynamics by invoking the
collapse
postulate.  According to this postulate, the state vector $\ket{\psi(t)}$, representing
a composite interacting `measured' and `measuring' system, stochastically collapses,
at some time $t'$ during their interaction, into one of $\ket{\psi(t')}$'s components in the
interaction basis.  The trouble is that this is not a way out
unless one can specify the physical conditions necessary and sufficient for a measurement
interaction to occur; for surely `measurement' is too
ambiguous a concept to be taken as primitive
 in a fundamental physical theory.

Collapse theories are designed to cure this defect in the collapse postulate.  They specify the
precise physical
conditions under which
collapses are more or less likely to occur, without treating `measurement' interactions as different
from other interactions contemplated by quantum theory.
Our main focus shall be on
\emph{wavefunction} collapse theories.  These are theories in which the representation of a state
vector $\ket{\psi(t)}$ as a function $\psi(t, \mathbf{r}_{1},\ldots, \mathbf{r}_{N})$ on the
configuration space of the system is taken to be
fundamental, and a collapse
increases the concentration of the amplitude of
$\psi(t, \mathbf{r}_{1},\ldots, \mathbf{r}_{N})$
 in some region of configuration space.

Collapse, so construed, faces two obstacles.
First, the smaller the region in which a wavefunction's amplitude is concentrated by a collapse, the
higher the collapsed state's dispersion will be in momentum space (by 
the uncertainty relation),
and hence the more energy
the system can possess after collapse.  So wavefunction collapses had better not make a
macroscopic system's wavefunction too narrow, otherwise the system could
spontaneously heat up in an observable way.
 Secondly, in order to be empirically adequate, wavefunction collapse theories 
 need to predict
that collapses of a microscopic system's wavefunction rarely occur, 
because the standard
Schr\"{o}dinger evolution of a microscopic system is overwhelmingly confirmed through
interference experiments.   But it is a well-known feature of the 
Schr\"{o}dinger equation that it prevents the wavefunction of a closed 
system of particles from ever developing a support confined to a bounded 
region of their configuration 
space (except at isolated instants of time)!\footnote{And this is not 
simply an artifact of nonrelativistic quantum mechanics.  
Restricting to the positive energy solutions of the Dirac equation, 
they all have infinite support (Thaller [1992]).  In fact, this is a 
consequence of the following much more general result. If the Hamiltonian 
generating the time evolution of a free relativistic 
particle has a spectrum bounded from below, and the particle is 
localized to a bounded region at $t=0$, then there is a nonzero 
probability of finding the particle \emph{arbitrarily far away} at 
\emph{any} later 
time (Hegerfeldt [1995]).}    
Now if a system's wavefunction cannot be made arbitrarily narrow by a collapse, 
and if it
can never be completely concentrated in a bounded region---i.e., if 
a system's wavefunction must virtually always possess `tails'
going off to infinity---then the standard semantics will block
the
attribution of a determinate location to each particle in the system, as well as to the position of
the system as a whole.  For a macroscopic system, it would then appear 
that the wavefunction collapse
theorist has little hope of finally putting the measurement problem to rest.  How one should deal
with
this problem, known as the  wavefunction \emph{tails problem}, is the subject of the present note.

In fact there is a standard solution to the
tails problem, which Albert and Loewer ([1996]) have recently argued for at length. The
solution is to weaken the eigenstate-eigenvalue link as regards the position of a particle by taking
a particle to be
located in some \emph{region} of space just in case its wavefunction is \emph{almost} an
eigenstate of being located in that region. Restricting attention to localizing the particle in a region
sidesteps the problem
that its wavefunction can never be infinitely narrow.  And taking the high probability of finding a
particle in a region to be sufficient for asserting that it actually is in
that region sidesteps the problem of infinite tails.
 Specifically, Albert and Loewer propose a weakened eigenstate-eigenvalue  link for position that
they
call PosR:
\begin{quote}
`Particle $x$ is in region $R$' if and only if the proportion of the total squared
amplitude of $x$'s wave function which is associated with points in $R$ is greater than or equal
to $1-p$.  (Albert and Loewer [1996], p. 87) \end{quote}
Albert and Loewer require that $p$ lie somewhere in the interval 
$(0,0.5)$ (for $p\geq 0.5$ would allow one to say that a particle lies in 
disjoint regions); otherwise, they argue that $p$ may be taken to
have any of the small continuum of values that can
underwrite the way we actually use the word `located' ([1996], p. 90).  The obvious generalization of
PosR to a multi-particle system would be (where $\times$ denotes 
Cartesian product):
\begin{quote}
`Particle $x$ lies in region $R_{x}$ \emph{and} $y$ lies in $R_{y}$ 
\emph{and} $z$ lies in $R_{z}$ \emph{and} $\ldots$'
if and only if the proportion of the total squared
amplitude of $\psi(t, \mathbf{r}_{1},\ldots, \mathbf{r}_{N})$ that 
is associated with points in $R_{x}\times R_{y}\times 
R_{z}\times\cdots$ is greater than or equal to $1-p$.
\end{quote}
We shall call this generalization of PosR, which Albert and Loewer do not explicitly endorse, the
\emph{fuzzy
link}.

At first glace it would seem that the fuzzy link, with a suitable value for $p$ selected, promises to
yield an unproblematic interpretation of
collapse theories, and allow them to fulfill their goal of representing everyday
macroscopic objects, like cats, as possessing reasonably
well-defined locations.  However, Peter Lewis ([1997]) has recently argued that the `spontaneous
localization' wavefunction collapse theory of GRW (Ghirardi, Rimini, and 
Weber [1986]), in
virtue of its need to rely on the fuzzy link to solve the tails problem, has the unacceptable
consequence that arithmetic does not apply to ordinary macroscopic objects.  We believe that
what Lewis succeeds in showing is that the GRW theory, interpreted in terms of the fuzzy link,
sometimes entails a failure of \emph{conjunction introduction}; that is, it entails that there can be
certain physical situations where a proposition $A_{1}$ is true, $A_{2}$ is true, $\ldots$\ , $A_{n}$ is true,
yet the conjunction $A_{1}\wedge A_{2}\wedge\ldots\wedge A_{n}$ (or $(\forall i)A_{i}$) is false. 
However,
we shall show that the GRW theory \emph{itself} guarantees that conjunction introduction can
\emph{never be experimentally falsified}, leaving the applicability of arithmetic to macroscopic
objects intact.

In the next section, $2$, we briefly review the essentials of the GRW theory needed for Lewis'
argument, and then spell out the
argument itself.  In section $3$, we show why the response to Lewis' argument recently given by
Ghirardi and
Bassi ([1999]) is unsuccessful, and consider other possible responses.  Section $4$ contains our demonstration that
the failure of arithmetic for macroscopic objects that Lewis alleges
(what we prefer to call a failure of conjunction introduction) can \emph{never}
become manifest in a world governed by GRW wavefunction collapses.  In our final
section, 5, we briefly discuss the larger issue: What epistemic stance should be taken towards
interpretations of quantum theory that are forced to posit `anomalies' which are never made
manifest?

\section{Lewis's Counting Anomaly}

We begin by recalling the ingredients of the GRW theory that are important 
to assessing Lewis' ([1997])
argument and Ghirardi and
Bassi's ([1999]) reply.

According to the GRW
theory, the quantum state of
an $N$-particle system evolves in accordance with Schr\"{o}dinger's equation \emph{except}
when a `hit' occurs on one of the particles in the system.
When
a `hit' occurs on the $i$th
particle, the total wave function $\psi(t, \mathbf{r})$
 for the system (abbreviating $(\mathbf{r}_{1},\ldots,
 \mathbf{r}_{N})$
 by $\mathbf{r}$)
instantaneously collapses to:
\begin{equation}
     \psi'(t, \mathbf{r})  =
     \frac{j(\mathbf{x}-\mathbf{r}_{i})\psi(t,
\mathbf{r})}{R_{i}(\mathbf{x})}.   
     \end{equation}
The jump factor $j$ effecting the hit
 is taken to be a normalized Gaussian of relatively narrow width $10^{-5}$ cm, and a hit on the
$i$th particle is posited to occur with probability $10^{-15}$ per 
second, for any $i$.
($R_{i}(\mathbf{x})$ is simply a renormalization factor, equal to $\int |j\psi| ^{2}
d^{3N}\mathbf{r}$.) 
The hit center $\mathbf{x}$, specified in $j$'s first argument, is randomly chosen
with probability
distribution $|R_{i}(\mathbf{x})|^{2} d^{3}\mathbf{x}$.  This     ensures that the localization
of the probability distribution for the $i$th particle to the region surrounding  a point
$\mathbf{x}$ in space occurs with
the probability given by the standard quantum-mechanical
Born rule.
For the quantum state of a \emph{microscopic}
system---a system with much less than $10^{15}$ particles---the particles in the system will
almost never be hit, so their total wavefunction will almost never collapse but, rather, will just
evolve in accordance with Schr\"{o}dinger evolution.
On the other hand, the
GRW theory ensures that the total wavefunction of a \emph{macroscopic} system---a system
with
more than $10^{15}$ particles---will collapse very rapidly.

There are some important features of GRW collapse that need to be kept in mind for what follows.
 Suppose an ordinary macroscopic object, like a marble, is in a superposition of different states,
where each state corresponds to the marble being well-localized to a region of space, but the
regions, which we denote by $L$ and $R$,
are widely separated:
\begin{equation}  \label{eq:in}
     \psi(t, \mathbf{r})  =
     c_{L}\ \psi_{L}(t, \mathbf{r})+
     c_{R}\ \psi_{R}(t, \mathbf{r}).
\end{equation}
 If one of the particles in the marble is
 hit (which is virtually inevitable, given the number of particles in the marble),  $\psi(t,
\mathbf{r})$
 will most likely collapse to a state which is
similar to the state of one of the two terms appearing in (\ref{eq:in}). This occurs because the
position of each particle in the
marble
is correlated to the position of every other particle, which ensures that when one particle is hit,
it is \emph{as if} the same jump factor were hitting every particle. Thus, if just one particle $i$
is hit with a jump factor $j(\mathbf{x}-\mathbf{r}_{i})$, and if $\mathbf{x}\in L$ (which will
occur with probability $\approx |c_{L}|^{2}$), then the result will be:
\begin{equation} \label{eq:out}
     \psi'(t, \mathbf{r})  =
     c_{L}\ j(\mathbf{x}-\mathbf{r}_{i})\psi_{L}(t, 
     \mathbf{r})/R_{i}(\mathbf{x})+  c_{R}\
j(\mathbf{x}-\mathbf{r}_{i})\psi_{R}(t, \mathbf{r})/R_{i}(\mathbf{x}). \end{equation}
Since the region in which $\psi_{R}$ is large is, by hypothesis, a region where $j$ is vanishingly
small, the norm of the second term is now very small relative to the first.  Thus a \emph{single}
hit has precipitated an effective collapse of the \emph{total} wavefunction $\psi(t, \mathbf{r})$
 onto its first term.  The collapse is only
 \emph{effective} since the second term of  (\ref{eq:out}) is never zero,  and will form part of the
residual `tails' of the wavefunction $\psi'(t, \mathbf{r})$ in
 configuration space\footnote{Our use of the phrase `effective 
 collapse' in this context should be sharply distinguished 
 from the way that phrase is sometimes employed in the context of 
 no-collapse interpretations to refer to the fact that---due to 
 `environmental decoherence'---the state of a macrosystem can be treated 
 \emph{for all practical purposes as if} it were no longer involved 
 in a superposition of macroscopically distinguishable 
 states.}.
  By contrast, suppose we have a pair of
non-interacting marbles in an unentangled product state:
\begin{equation}
     \psi(t, \mathbf{r},\mathbf{r}')  =
     \psi_{1}(t, \mathbf{r})
     \psi_{2}(t, \mathbf{r}').
\end{equation}
 In this case, any number of hits on the particles in one of the marbles will leave
 the other marble's wavefunction unchanged.  For example,
 a hit on the $i$th particle in
marble 1 will just produce:
\begin{equation} \label{eq:four}
     \psi'(t, \mathbf{r},\mathbf{r}')
     =  \frac{j(\mathbf{x}-\mathbf{r}_{i})
     \psi_{1}(t, \mathbf{r})}{R_{i}(\mathbf{x})}
     \psi_{2}(t, \mathbf{r}')
     \end{equation}      
(where $R_{i}(\mathbf{x})=\int |j\psi_{1}| ^{2} d^{3N}\mathbf{r}$). It follows that in the
absence of entanglement between the marbles, they will each be subject to independent GRW
collapse processes that will preserve the product character of their total wavefunction for as long
as they fail to interact.

     We turn now to Lewis's ([1997]) argument.
       He first considers a marble that can be either in or out of a box.  The eigenstate of the
marble being in the box is $\inket$, while the eigenstate of the marble being outside the box is
$\outket$. Suppose the  marble starts out in the state
\begin{equation}
     \frac{1}{\sqrt{2}}(\inket + \outket)   
     \end{equation}
(which would correspond to a wavefunction $\psi(t, \mathbf{r})$ with both
$\int_{\mbox{in}}|\psi| ^{2} d^{3N}\mathbf{r}$ and
$\int _{\mbox{out}} |\psi| ^{2} d^{3N} \mathbf{r}$ equal to $1/2$).  Lewis observes
that GRW's collapse dynamics will almost
always leave
the system in
either a state like
\begin{equation} \label{eq:six}
     a\inket + b\outket,
     \end{equation}
     or a state like
     \begin{equation}
     a\outket + b\inket,
     \end{equation}
 where $1>|a|^{2}\gg |b|^{2}>0$. Applying the
eigenstate-eigenvalue link to either collapsed state, the marble is neither in nor out of the
box.  But according to the \emph{fuzzy} link for
$|b|^{2}\leq p$,
the marble is either determinately in or determinately out of the box. Moreover, if we suppose the
post-collapse state of the marble is in fact (\ref{eq:six}), so that
according to the fuzzy link the marble is in
the box, then when one measures the location of such a marble, one could obtain the result that
the marble is out of the box. The probability of this happening, $|b|^{2}$, is extremely low, but
according to the GRW theory it
could happen.  Still, there is nothing contradictory in this state of affairs. One simply has to accept
that \emph{very} rarely a measurement of a marble's location will cause it to jump to a location
disjoint from the location it had, according to the fuzzy link, prior to 
the measurement. Indeed, even in the absence of measurement, the 
marble could well 
make such a jump spontaneously.  Lewis's
counting anomaly can be seen as an attempt to magnify this 
unlikely anomaly to the point of absurdity.

Thus, Lewis next considers a system of $n$ non-interacting marbles, each of which is in a state
like (\ref{eq:six}):
\begin{equation}
     \ket{\psi}_{\mbox{all}}=  (a\inket_{1} + b\outket_{1})\otimes    (a\inket_{2} +
b\outket_{2})\otimes\cdots\otimes
     (a\inket_{n} + b\outket_{n}).
     \end{equation}
Lewis takes each $\inket_{i}$ state to refer to localization within a 
single box which is the same for all the marbles.  He also needs to 
assume that the marbles are noninteracting, which can be ensured by 
making the dimensions of the box sufficiently large\footnote{Alternatively, 
one could assume that each state $\inket_{i}$ refers to a separate 
box, and that the $n$ boxes, one for each marble, are pairwise widely 
separated in space.}.  When one measures the location of each marble in turn, the probability that all $n$ marbles will be
found to be in the box is
\begin{equation}
     |\bra{\psi}_{\mbox{all}}\
     \inket_{1}\otimes\inket_{2}\otimes\cdots\otimes\inket_{n}|^{2}   =  
     |a^{n}|^{2}=|a|^{2n}.   
 \end{equation}
Since $|a|^{2}<1$ (notwithstanding that $|a|^{2}$ could be quite 
close to $1$), $|a|^{2n}\ll 1$ for sufficiently large $n$, which makes it highly
unlikely that all the marbles will be
found in the box. Lewis concludes from this that the state $\ket{\psi}_{\mbox{all}}$
`cannot be one in which all $n$ marbles are in the box, since there is almost no chance that if one
looks one will find them there' ([1997], p. 318).
In other words, by applying the fuzzy link for $|a|^{2n} \leq p$ to $\ket{\psi}_{\mbox{all}}$,
one
obtains
the result that not all the marbles are in the box. And this seems to contradict the results one
obtains when one applies the fuzzy link on a marble-by-marble basis, where one gets the
results that marble 1 is in the box, marble 2 is in the box, and so on 
through marble $n$.\footnote{Note the similarity 
between this result and the `lottery paradox'.  
For any particular ticket holder 
(in a lottery with a sufficiently 
large number of tickets) one is inclined to infer, from the ticket 
holder's high probability 
of losing, that they will in fact lose (cf. `this particular marble is in the box').  Yet, if
the same inference is made for each ticket holder in turn, we would 
apparently arrive at the absurd conclusion that we are certain that no one 
will win the lottery (cf. `all the marbles are in the box').  
But note well: while we would be wholly within our rights to draw back from the 
inference from high probability to certainty that generates 
the lottery paradox, exercising 
the same freedom against Lewis's counting paradox would mean 
rejecting the fuzzy link and, consequently, send the 
wavefunction collapse theorist right back to the tails problem.}    

     Actually, as Lewis points out ([1997], p. 321), the contradiction only holds if one assumes
\emph{the enumeration principle}: `if marble 1 is in the box and marble 2 is in the box and so on
through marble $n$, then all $n$ marbles are in the box'. This principle is but a special case of
conjunction introduction: if $A_{1}$ is true, $A_{2}$ is true, and so on through $A_{n}$,
then $A_{1}\wedge A_{2}\wedge\ldots\wedge A_{n}$ (that is, $(\forall i) A_{i}$) is true.
Thus the fuzzy
link, the GRW theory, and conjunction introduction jointly entail a contradiction.
The moral Lewis draws from this contradiction is that \emph{no-collapse}
interpretations of quantum theory not subject to the tails problem, such 
as Bohm's ([1952]) theory,
are to be
preferred over wavefunction collapse theories like GRW's.  We believe this conclusion is far too
quick, so our next task
will be to examine some other possible routes around the contradiction. 
\section{Can the Counting Anomaly be Avoided?}    

Recently Ghirardi and Bassi
([1999]) have claimed that Lewis's contradiction between the fuzzy link and and the enumeration
principle (or conjunction introduction)  in fact \emph{fails to arise} in the GRW theory.
They consider the system in state $\ket{\psi}_{\mbox{all}}$, and point out
that it can be rewritten as the following superposition of $2^{n}$ macroscopic states: \begin{eqnarray} 
\ket{\psi}_{\mbox{all}} =
a^{n}\inket_{1}\otimes\inket_{2}\cdots\inket_{n}+
a^{n-1}b\outket_{1}\otimes\inket_{2}\cdots\inket_{n}
\nonumber  \\ \label{eq:atbest}
+ a^{n-1}b\inket_{1}\otimes\outket_{2}\cdots\inket_{n}
+\cdots + a^{n-1}b\inket_{1}\otimes\inket_{2}\cdots
\outket_{n} \\
+a^{n-2}b^{2}\outket_{1}\otimes\outket_{2}\cdots\inket_{n}
+\cdots + b^{n}\outket_{1}\otimes\outket_{2}\cdots
\outket_{n}. \nonumber
\end{eqnarray}
They then write:
\begin{quote}
The marbles are macroscopic objects, and, as such, they contain a number of particles of the order
of Avogadro's number. But it is the
most fundamental physical characteristic of the GRW theory that it forbids the persistence of
superpositions of states of this kind. In particular for the case 
under consideration the precise
GRW dynamics will lead in about one millionth of a second to the suppression of the
superposition and the ``spontaneous reduction'' of the state [i.e. $\ket{\psi}_{\mbox{all}}$] to
one of its terms (with the probability attached to it by
its specific coefficient).  (Ghirardi and Bassi [1999], p. 708)
\end{quote}
Ghirardi and Bassi go on to argue ([1999], Section 4) that even if we assume $n$ 
to be so large that all the mass of the universe is used to constitute 
the $n$ marbles, $\ket{\psi}_{\mbox{all}}$ is still overwhelmingly likely to 
GRW collapse to the \emph{first} term in $(\ref{eq:atbest})$ given 
how close $|a|^{2}$ will in fact be to 1 for an object the size of a 
marble undergoing GRW collapse.   But their
main conclusion, with reference to  
$\ket{\psi}_{\mbox{all}}$, is that 
`it is meaningless to make any statement about the location of the marbles in such
states simply because they cannot persist for ``more than a split 
second''\ ' ([1999], p. 709).   
     
     Let us suppose that Ghirardi and Bassi are 
     correct that GRW collapse will rapidly produce a reduction of the state 
     $\ket{\psi}_{\mbox{all}}$ to one of the terms in 
     (\ref{eq:atbest}) (setting aside, for the moment, to 
     \emph{which} term $\ket{\psi}_{\mbox{all}}$ is most likely to 
     collapse).  Of course, if such a reduction were to occur, it would 
     have to happen in such a way that the squared modulus of the 
     coefficient of one of the terms in (\ref{eq:atbest}) became large, and the other 
     coefficients became small; for a \emph{perfect} collapse to one of 
     (\ref{eq:atbest})'s terms would leave the individual 
     wavefunctions of the marbles without tails.  Also recall from 
     Section 2 that, 
since the noninteracting marbles begin in the product state 
$\ket{\psi}_{\mbox{all}}$ (which \emph{is} a product state, notwithstanding 
the fact that $\ket{\psi}_{\mbox{all}}$ may be rewritten as in 
(\ref{eq:atbest})), the marbles' final 
\emph{effectively} collapsed state must \emph{again} be a product 
state.  So no matter what GRW collapses occur in state (\ref{eq:atbest}),  the final state 
must have the form:
\begin{equation} \label{eq:ugh}
     \ket{\psi'}_{\mbox{all}}=  (a_{1}\inket_{1} + b_{1}\outket_{1})\otimes    (a_{2}\inket_{2} +
b_{2}\outket_{2})\otimes\cdots\otimes
     (a_{n}\inket_{n} + b_{n}\outket_{n})
     \end{equation}
where all the $a_{i}$'s and $b_{i}$'s are nonvanishing.  
Ghirardi and Bassi's claim, on behalf of $\ket{\psi'}_{\mbox{all}}$, 
must then be that when we 
re-expand $\ket{\psi'}_{\mbox{all}}$, as $\ket{\psi}_{\mbox{all}}$ 
was expanded in (\ref{eq:atbest}) above, the absolute square of 
one of 
the terms' coefficients (there will again be $2^{n}$ terms) will now be $\geq 1-p$.  
However, by hypothesis we started out 
in a state  $\ket{\psi}_{\mbox{all}}$ wherein all the squares of the coefficients in (\ref{eq:atbest}) 
were bounded above by $|a|^{2n}$, which itself was supposed to be much 
less than $1$.  So the only way for the 
marbles to end up as Ghirardi and Bassi claim 
is that a large number of the marbles in $\ket{\psi}_{\mbox{all}}$, in 
a very short time, have either their $a$ or $b$ coefficients  driven by 
GRW collapse processes 
closer (in absolute square) to 
1 than $a$'s value was.  That is, the only way to make sense of 
Ghirardi and Bassi's claim (consistent with Lewis's standing 
assumptions that the marbles form an isolated system and do not 
interact, assumptions that \emph{entail} that the marbles will be 
subject to independent GRW collapses that preserve the product 
character of their total state) is to suppose that in the final state 
$\ket{\psi'}_{\mbox{all}}$ various coefficients, say $a_{1}$, $b_{2}$, $a_{3}$, etc. will 
have their absolute squares much closer to $1$ than $a$'s was, 
yielding a value $|a_{1}|^{2}|b_{2}|^{2}|a_{3}|^{2}\cdots\geq 1-p$ that 
restores 
consistency with the enumeration principle.  

But herein lies the rub.  For the purposes of his argument, Lewis 
supposed that the initial product state of the 
marbles was \emph{already} one in which $|a|^{2}$ had been driven 
by GRW collapse as close to $1$ as it can be (a perfectly legitimate 
assumption, since if there were no upper limit on $|a|^{2}$, there would be no tails problem to 
begin with).  He was then free to choose a 
sufficiently large value of $n$ with which to run his argument.  It follows 
that the collapse scenario Ghirardi and Bassi envisage, in which some 
of the 
$|a|^{2}$'s get \emph{still closer} to $1$, has already been taken into account by Lewis's 
argument, and cannot supply a basis from which to launch a criticism 
of that argument.  Here is another way to see the point.  Suppose we 
grant that $\ket{\psi}_{\mbox{all}}$ very quickly 
evolves to a state of form $\ket{\psi'}_{\mbox{all}}$.  Then Lewis would 
still be free to exploit this latter state as the starting point for his 
argument.  He could, first, drop from consideration those marbles 
where $|a_{i}|^{2}<1-p$.  For the marbles that remain, they will be in a product 
state $\ket{\psi''}_{\mbox{all}}$ again of form (\ref{eq:ugh}) where, now, 
\emph{all} 
the $a_{i}$'s have absolute squares within $p$ of 1 (and thus each 
individual marble will be in the box, according to the fuzzy link).  
Lewis could, then, 
simply consider sufficiently many marbles 
in a product state formed from sufficiently many tensor products of 
$\ket{\psi''}_{\mbox{all}}$ with itself in order to guarantee, yet 
again, that the 
probability that they are \emph{all} in the box is less than or equal 
to $p$!    

This leads us to Ghirardi and Bassi's other criticism of Lewis's 
argument.  In effect, they challenge Lewis's application of 
the multi-particle fuzzy link to $\ket{\psi}_{\mbox{all}}$ (or, if you 
prefer, to 
tensor products of $\ket{\psi''}_{\mbox{all}}$ with itself) by 
questioning whether it would actually be possible to produce enough 
marbles so that the probability that they are all in the box in
 a state like $\ket{\psi}_{\mbox{all}}$  is 
less than or equal to some small number $p$.  
They calculate that if each marble possesses a mass 
of about 1 gram, which puts precise limits on the value of 
$a$, and if we allow ourselves the mass of the entire 
universe ($\approx 10^{53}$ grams) with which to constitute the marbles, 
which sets a limit on the value of $n$, 
then the
overwhelmingly most 
likely configuration of the marbles in state $\ket{\psi}_{\mbox{all}}$ 
will, in actual fact, \emph{still} be the one
 where they 
are all in the box.  In other words, under the given assumptions about 
marbles and our universe, it turns out that $|a|^{2}\geq 
1-p$ for any reasonably small value for $p$.  We see no reason to 
doubt this. However, it is at best a contingent fact 
about our world, since no matter how close to $1$ GRW collapses will make 
$|a|^{2}$ for a macroscopic object like a marble, one could 
always imagine a sufficiently massive universe 
where $n$ is 
large enough that $|a|^{2n}$ is close to $0$ and the 
contradiction between the fuzzy link, conjunction introduction, and 
the GRW theory remains.  We believe a stronger response to Lewis's counting 
anomaly would be one in which the force of the anomaly is 
muted for reasons internal to the GRW theory itself.  
Moreover, although Lewis clearly wants the 
anomaly to obtain for macroscopic objects, there would still be 
something puzzling about its obtaining for microscopic objects like 
particles.  Yet if we replace each marble with a particle, which will 
be hit only very rarely, there \emph{is no} reason internal to the GRW theory 
why $|a|^{2}$ should have to be 
so close to $1$ that $|a|^{2n}$ cannot be close to $0$ for 
a large number $n$ of particles in \emph{our} universe.       

Where do these shortcomings in Ghirardi and Bassi's 
reasoning leave their claim that  $\ket{\psi}_{\mbox{all}}$ will be  
`transformed immediately into a perfectly reasonable (from the point 
of view of the enumeration principle) state' ([1999], p. 709)?  Let us 
see.    We can either suppose that the marbles are microscopic particles, or that 
the universe is sufficiently massive; we shall persist in telling the 
story using marbles in the state $\ket{\psi}_{\mbox{all}}$ (and the 
product of $\ket{\psi''}_{\mbox{all}}$ with itself enough times would 
do equally well).  Since the marbles are
non-interacting by hypothesis, when one of the particles in a
marble is hit, the states of the
other
marbles will not be affected. Obviously if no hits produce marbles 
jumping out of the box, then the failure of the enumeration principle will 
persist. So suppose, instead, that a collapse occurs in such a way that
marble 1 jumps out of the box.  This is a very rare occurrence, but with enough marbles, one of them
 is
bound to make the jump, and we lose no generality in assuming it is marble 1.  The state of the
system will then be \begin{equation} \label{eq:fourteen}
\ket{\psi}_{\mbox{out}_{1}}=  (c\inket_{1} + d\outket_{1})\otimes     (a\inket_{2} +
b\outket_{2})\otimes\cdots\otimes
     (a\inket_{n} + b\outket_{n}),
     \end{equation}
     where $|d|^{2}\approx |a|^{2}$ (supposing, once more, that the value 
     $|a|^{2}$ is about
     as 
     close to $1$ as GRW collapse can achieve).  Now let $A_{1}$ denote  the proposition that
`marble 1
is in the box', $A_{2}$ that `marble 2 is in the
box', and so on. According to the fuzzy link applied to each marble in state
$\ket{\psi}_{\mbox{out}_{1}}$,
it is true that marble 1 is \emph{not} in the box,
that marble 2 is in the box, 3 is in the box, and so on through marble $n$. By conjunction
introduction, then,
$\neg A_{1}\wedge A_{2}\wedge\cdots\wedge A_{n}$ is true.
Yet with the fuzzy link applied to all $n$ marbles, this conjunction is \emph{false}, since
\begin{equation}
     |\bra{\psi}_{\mbox{out}1}\
     \outket_{1}\otimes\inket_{2}\otimes\cdots\otimes\inket_{n}|^{2}  =  
     |da^{n-1}|^{2} = |d|^{2}|a|^{2(n-1)},  
\end{equation}
and, by hypothesis, $|d|^{2}|a|^{2(n-1)}\approx |a|^{2n} \leq p$ for sufficiently large $n$. Thus, the contradiction still obtains and the
initial failure of conjunction introduction will simply
be propagated via GRW evolution into the failure of another instance 
of conjunction introduction.  This remains true
no matter how many marbles make jumps or how 
rapidly GRW collapses occur.  For at the moment after any number of 
jumps have occurred, there will always remain some conjunction relative to 
which conjunction introduction fails.  The conclusion, we think, is inevitable: 
GRW collapse evolution in fact \emph{cannot} suppress the failure of 
conjunction introduction in a sufficiently large isolated system of non-interacting 
marbles that evolve from a product 
state.    

There are at least two other possible strategies one might adopt to avoid Lewis's counting
anomaly in
the GRW theory.  Both strategies
involve modifying the fuzzy link.

One route around Lewis's counting anomaly might be to employ
different values for the fuzzy link's $p$: one value, $p$, when the fuzzy link is applied to an
individual marble's probability distribution, and another value, $p_{\mbox{all}}$, when the link is applied
to their total distribution as determined by $\ket{\psi}_{\mbox{all}}$.  
For any range of values in the interval $(0,0.5)$ that one believes it is
necessary to insist upon in order to underwrite our
uses of the term `located', one could always make sure 
both $p$ and $p_{\mbox{all}}$ are chosen from within one's preferred range in such a way
that $(1-p)^{n}\geq 1-p_{\mbox{all}}$.  If, then, 
each of the marbles has at least $1-p$ of its probability
concentrated within the box, i.e. if $|a|^{2}\geq 1-p$, then each 
will get counted as actually in the box, and the \emph{conjunction} of all those assertions will have
the minimum necessary probability of $1-p_{\mbox{all}}$ for \emph{it} 
to be counted 
as true as well!  On the other hand, if $|a|^{2}< 1-p$, so that no 
single marble is in the box, or 
if $|a|^{2n}<1-p_{\mbox{all}}$, so that they are not all in the box, 
then the issue of a failure of conjunction 
introduction does not arise.     
Unfortunately, this clever strategy simply resurrects 
the tails problem.  To guard against failures of conjunction 
introduction, one would always need to choose $p$ so that $(1-p)^{n}\geq 
1-p_{\mbox{all}}>0.5$.  But this means that as $n$ 
grows large, $p$ must be chosen closer and closer to zero
 no matter what value is assumed for $p_{\mbox{all}}$.  
 Moreover, as we have seen, there will be 
 always some 
 value, $|a|^{2}$, close to $1$ such that GRW collapses cannot 
 localize the marbles to the box with a higher probability than 
 $|a|^{2}$ (and this would also be true were the marbles replaced by 
 particles).   Therefore, we could consider a sufficiently
  large value of $n$
such that one is forced by the inequality $(1-p)^{n}>0.5$ to choose a 
$p$ satisfying $|a|^{2}<1-p$.  
In that case, though no failure of conjunction introduction ensues, 
the fuzzy link will dictate that
no marble will be in the box.   So \emph{no matter how well} GRW collapses 
can concentrate the marbles' wavefunctions in the box, we would be 
forced to conclude that no marble can be in the box, returning 
us right back to the tails problem.  

Second, one might argue that the correct way to resolve the tails problem is to endorse just
PosR---the fuzzy link applied \emph{only} on an individual particle basis.  One could then take
facts about the joint positions of systems of particles, like marbles and collections of marbles, to
supervene directly on facts about the positions of individual
particles, rather than on their total wavefunction.
Albert and Loewer themselves only formulate the fuzzy link for individual particles, and their
willingness to suppose that `the value of the cat's aliveness is determined by the positions of the
particles that make it up' ([1996], p. 86) suggests that they might prefer this strategy.  Certainly this
strategy would allow one to assert that all
the marbles are in the box, notwithstanding their total state $\ket{\psi}_{\mbox{all}}$. 
However, since $\ket{\psi}_{\mbox{all}}$ is almost an
eigenstate of `all the marbles are in the box' being false, this strategy would require that the
wavefunction collapse theorist not simply weaken the
eigenstate-eigenvalue link between truth and probability
1, but sever this link entirely.
And if one is willing to entertain the thought that events in a quantum world can happen without
being mandated or made
overwhelmingly likely by the
wavefunction, then it is no longer clear why one should need to solve the measurement problem
by
collapsing wavefunctions!  Another reason not to restrict to PosR alone is that it seems arbitrary
to apply a semantic rule for quantum states to a single-particle system, but not to a
multi-particle system.  Indeed, to the extent that one supposes there to be a plausible intuitive
connection between
an event's having high probability according to a theory, and the event actually occurring, one is
hard-pressed to resist the intuition in the multi-particle case.  Finally, it is not clear how one could
even make the distinction between PosR and the (full) fuzzy link in more sophisticated theories,
like the
`continuous spontaneous localization' theory of Ghirardi, Grassi, and Pearle 
([1990]), where talk of
particles is replaced by talk of systems in near eigenstates of local mass density.

Even if one does not regard the above considerations as decisive against modifying the fuzzy link,
neither Lewis nor Ghirardi and Bassi question this link, and we shall argue in the next section that the
price of sometimes abandoning
conjunction introduction is not near as high as Lewis portrays.  But lest one think that conjunction
introduction is an analytic truth, it is worth pointing out that abandoning analogous principles in
the context of the interpretation of quantum theory is not unprecedented.  In the quantum logic
that Kochen
and Specker ([1967]) advocate, conjunctions of quantum-mechanically incompatible propositions are
not syntactically well-formed, and in Bell's ([1986]) quantum logic, 
conjunction introduction is an invalid inference rule.  Moreover, as Clifton 
([1996],
p. 386) points out, various modal interpretations deny \emph{property composition}, a principle
which roughly says that if system $S_{1}$ has some property and system $S_{2}$ has some
property, then system $S_{1}+S_{2}$ has the corresponding joint property.  Property
composition can be seen as a version of conjunction introduction, and note that the `conjuncts', in
this case, are \emph{compatible}.

But, it will be argued: so much the worse for quantum logics and modal interpretations!  However,
as Albert and Loewer ([1996], p. 87) point out, even before multi-particle systems are considered, the
GRW theory, interpreted via
the PosR rule, violates the principle of \emph{property intersection}: that if particle $x$ lies in
region $\Delta$, and $x$
lies in
$\Delta'$, then $x$ must lie in $\Delta\cap\Delta'$.  The
violation of this principle is an easy consequence of PosR's identification of `$x$ lying in $\Delta$'
with
`$x$ having a probability of at least 1-$p$ of being found in $\Delta$'. And it is at least arguable
that once property intersection fails, so must conjunction introduction---at least if we want `$x$
lying in $\Delta$' to mean `no part of $x$ lies
outside of $\Delta$'.  For if no part of $x$ lies outside of $\Delta$, and no part lies outside of
$\Delta'$, then, in particular, no part of $x$ lies inside $\Delta$ but \emph{outside} $\Delta'$, nor
does any part of $x$ lie inside $\Delta'$ but \emph{outside} $\Delta$.  By conjunction
introduction, then, it follows that no part of $x$ lies outside of $\Delta\cap\Delta'$, i.e., that $x$
lies in $\Delta\cap\Delta'$, and we have derived property intersection.  
(Clifton ([1996], pp. 381-2)
gives this same
argument to highlight the import of property intersection's violation in Healey's modal
interpretation.)

We are not suggesting by these remarks that anomalies, such as the failure of conjunction
introduction, should simply be swept under the carpet. Indeed, Albert and
Loewer take a major part of their task to be to 
provide reasons for thinking that with a sufficiently small value chosen for $p$ in PosR,
violations of property intersection `aren't going to be worth bothering 
about' ([1996], p. 89).  Our
final task will be to make the same kind of point in relation to Lewis' multiple-particle failure of
conjunction introduction (though, as we have seen above, 
playing with the value of $p$ will be of no help).

\section{Is the Counting Anomaly Ever Manifest?}

     Recall that Lewis presents his argument as demonstrating a failure of the  enumeration
principle.  He gives the following argument for   why we
should not give up this principle:
\begin{quote}
\emph{If we want to maintain that the enumeration principle breaks down at some point, then, we
must maintain that the process of counting marbles breaks down}---that counting cannot be
applied to sufficiently large systems of marbles. Since counting is the foundation of arithmetic, this
is tantamount to saying that arithmetic does not apply to sufficiently large systems of marbles
[italics ours].  (Lewis [1997], p. 321)
\end{quote}
Lewis then argues, and we agree, that the cost of holding that arithmetic does not
apply to large systems of marbles is too high.
But we do \emph{not} agree that
 Lewis has established the conditional that we italicized above.  The
 trouble is that Lewis fails to operationalize the process of counting  marbles by explicitly
modelling \emph{the process itself} in terms of collapsing GRW
 wavefunctions.   Indeed, by calling the failure at issue a failure of the enumeration
 principle, with all the operational connotations of the term `enumeration',   Lewis fails to keep the
 failure of conjunction introduction distinct from its empirical  falsification.  Moreover, taking the
objects at issue to be  \emph{macroscopic} marbles conveys the impression
 that the relevant failure of `enumeration'
 is somehow already manifest in state $\ket{\psi}_{\mbox{all}}$, but that is only the case if we
allow the enumerator the powers of Maxwell's demon. 
 Lewis shows signs of being aware of this concern, when
 he writes (just before the remarks quoted above):
 \begin{quote}
 To deny the enumeration principle, we must deny that if we put marble 1 in the box, and we put
marble 2 in the box, and so on through marble $n$, then there are $n$ marbles in the box.  \emph{This
process of putting marbles in a box is essentially one of counting the marbles} [italics ours].
\end{quote}
However, if marbles are `put into' the box one-by-one, presumably interactions between marble
placing devices and marbles will have to take place.  But then it is no longer clear how one could
thereby prepare the state
$\ket{\psi}_{\mbox{all}}$, which presupposes that the marbles are not entangled in any way with
each other or their environment. In any case, rather than seeking to make Lewis's `putting'
metaphor physically concrete, we
shall simply grant that $\ket{\psi}_{\mbox{all}}$ has been prepared 
by some means and, instead,
ask whether the marbles' (or particles') instantiation of a failure of conjunction introduction can ever
become manifest to a physical observer who takes it upon herself to tally 
them up.   Our answer is `No', and it will be
clear that our considerations also go through for a system in state
$\ket{\psi}_{\mbox{out}_{1}}$ and in any other of the states to which
$\ket{\psi}_{\mbox{all}}$ (or relevantly similar product states of 
micro- or macroscopic objects) can
evolve.

The most straightforward way to manifest a failure of conjunction introduction for a system in
state $\ket{\psi}_{\mbox{all}}$ would be to empirically
establish the truth of each of $A_{1}$, $A_{2}$, $\ldots$ , $A_{n}$, together with an
independent empirical test of the truth of $\neg(A_{1}\wedge A_{2}\wedge\cdots\wedge
A_{n})$.

An ideal measurement of whether marble 1 is in the box would correlate orthogonal states of a
macroscopic measuring apparatus to the $\inket$ and $\outket$ states of the marble.
Since this must be done for
all $n$ marbles, $n$ apparatuses must be used. The marbles/apparatuses system will evolve from
the state
\begin{equation}
     [(a\inket_{1} + b\outket_{1})\otimes\cdots\otimes
     (a\inket_{n} + b\outket_{n})]\otimes\ket{\mbox{ready}}_{M_{1}}   
\otimes\cdots\otimes \ket{\mbox{ready}}_{M_{n}}
     \end{equation}
     to the state
\begin{equation} \label{eq:fifteen}
     (a\inket_{1}\ket{\mbox{`in'}}_{M_{1}} +
     b\outket_{1}\ket{\mbox{`out'}}_{M_{1}})\otimes\cdots\otimes      
(a\inket_{n}\ket{\mbox{`in'}}_{M_{n}} +
     b\outket_{n}\ket{\mbox{`out'}}_{M_{n}}).
     \end{equation}
Since $|b|^{2}\leq p$, the fuzzy link dictates that each
apparatus records that its marble is in the box.
     This procedure does not yet qualify as a counting procedure, since    we have not yet
modelled an apparatus which records
     \emph{how many} marbles are in the box. One
might think that the information about each individual marble in the marble apparatuses, which
could be further correlated to different memory stores in an observer's brain, could
simply be combined `in thought' to get direct information about how many marbles are in
the box.  However, if we are not simply going to beg the question against verifying the
enumeration principle, acquiring information about the marble count must itself be modelled in the
GRW theory by a further interaction with
the marbles/apparatuses system or within the observer's brain.

     Let us turn, then, to a generic counting procedure that will establish how many marbles are in the
box. No doubt there are many ways to implement counting physically, but the general scheme will
need to involve a measurement on the marbles/apparatuses system that is the equivalent of asking the question:
`How many
marbles are in the box?'. Let $O$ be an observable with $n+1$ eigenvalues $o_{i}$, where the
$o_{i}$-eigenspace of the operator associated with $O$ is the subspace spanned by all the terms
in the superposition (\ref{eq:fifteen}) which have as coefficient $a^{i}b^{n-i}$. Thus, a system in
an $o_{i}$-eigenstate of
 $O$ is one where exactly $i$ of the $n$
marbles are in the box.
Consider a measurement of the observable $O$ on a system in state (\ref{eq:fifteen}).  After the
measurement, the system is in the state \begin{eqnarray} \label{eq:sixteen}
\ket{\psi}_{\mbox{count}} & =  &
a^{n}\ket{\phi}_{\mbox{out}_{0}}\ket{\mbox{`}O=n\mbox{'}}_{M}+
a^{n-1}b\ket{\phi}_{\mbox{out}_{1}}\ket{\mbox{`}O=n-1\mbox{'}}_{M} \\ \nonumber & &
+\cdots+
b^{n}\ket{\phi}_{\mbox{out}_{n}}\ket{\mbox{`}O=0\mbox{'}}_{M}, \end{eqnarray}
where
\begin{eqnarray} \nonumber
     \ket{\phi}_{\mbox{out}_{0}} & = &
     \inket_{1}\ket{\mbox{`in'}}_{M_{1}}\inket_{2}\ket{\mbox{`in'}}_{M_{2}}\cdots 
     \inket_{n}\ket{\mbox{`in'}}_{M_{n}}  \\ \nonumber
          \ket{\phi}_{\mbox{out}_{1}} & = &
     \ \ \outket_{1}\ket{\mbox{`out'}}_{M_{1}}
     \inket_{1}\ket{\mbox{`in'}}_{M_{2}}\cdots
     \inket_{n}\ket{\mbox{`in'}}_{M_{n}} \\ \nonumber
          &  &
     +\inket_{1}\ket{\mbox{`in'}}_{M_{1}}
     \outket_{1}\ket{\mbox{`out'}}_{M_{2}}\cdots
     \inket_{n}\ket{\mbox{`in'}}_{M_{n}} \\ \nonumber
     & & \vdots  \\ \nonumber
          &  &
     +\inket_{1}\ket{\mbox{`in'}}_{M_{1}}
     \inket_{1}\ket{\mbox{`in'}}_{M_{2}}\cdots
     \outket_{n}\ket{\mbox{`out'}}_{M_{n}} \\ \nonumber
     \vdots & & \\ \nonumber
     \ket{\phi}_{\mbox{out}_{n}} & = &
     \outket_{1}\ket{\mbox{`out'}}_{M_{1}}
     \outket_{2}\ket{\mbox{`out'}}_{M_{2}}\cdots
     \outket_{n}\ket{\mbox{`out'}}_{M_{n}}.
     \end{eqnarray}
     Since $|a|^{2n}\leq p$, by the fuzzy link it is not the case that all $n$ marbles are in the
box. However, each individual marble is still in the box, since for all $i$,
$|\bra{\psi}_{\mbox{count}}\inket_{i}|^{2}\geq 1-p$.
Thus, we have a violation of the enumeration principle and hence conjunction introduction.

But this does not mean that a failure of the rules of counting has now become manifest! The state
$\ket{\psi}_{\mbox{count}}$ is highly unstable given the GRW dynamics, since we see from
(\ref{eq:sixteen}) that  it is an \emph{entangled} superposition of states of
macroscopic systems, where the various
terms of (\ref{eq:sixteen}) markedly differ as to the location of the pointer on $M$'s dial that
registers the value of $O$. Thus, the GRW dynamics dictates that it is very likely that the total
system will effectively collapse onto one of the terms in
(\ref{eq:sixteen}), and that it will do so very quickly, given how many particles in
$M$, the $M_{i}$ apparatuses, and the marbles have the potential to be 
hit\footnote{What is important here is that the total 
number of particles involved in (\ref{eq:sixteen})'s entangled state is sufficiently 
macroscopic.  Thus, the macroscopic apparatuses $M_{i}$ and $M$ could 
be replaced by 
single particles, whose positions act as the measurement pointers, 
and GRW dynamics would \emph{still} guarantee effective collapse to one of 
the terms in (\ref{eq:sixteen}).  
(And if, further, the collection of marbles is replaced by a collection of 
particles sufficiently large to produce Lewis's counting anomaly, 
their number will also more than likely suffice to produce the same 
sort of effective collapse of (\ref{eq:sixteen}).)}. 
If the effective collapse is onto the state
     $\ket{\phi}_{\mbox{out}_{0}}\ket{\mbox{`}O=n\mbox{'}}_{M}$, then clearly no
failure of conjunction introduction becomes manifest, since the results of the various individual
apparatuses in that state are in
agreement with $M$. What if the system effectively collapses onto some other term of
(\ref{eq:sixteen}), such as
$\ket{\phi}_{\mbox{out}_{1}}\ket{\mbox{`}O=n-1\mbox{'}}_{M}$?  In this case, since
$\ket{\phi}_{\mbox{out}_{1}}$ is \emph{itself} an entangled state, since 
its terms (pairwise) differ as to
location of at least one of
the marbles, and since the $M_{i}$ apparatuses and marbles are 
macroscopic (or, if we are enumerating particles instead of marbles: since their 
number is extremely large), there will a further quick, effective
collapse to one of
$\ket{\phi}_{\mbox{out}_{1}}$'s
terms.  Suppose, for example, that the total state (effectively) ends up as \begin{equation}
\label{eq:seventeen}
     (\outket_{1}\ket{\mbox{`out'}}_{M_{1}}
     \inket_{2}\ket{\mbox{`in'}}_{M_{2}}\cdots
     \inket_{n}\ket{\mbox{`in'}}_{M_{n}})\ket{\mbox{`}O=n-1\mbox{'}}_{M}.  
\end{equation}
Then once again the results of the individual apparatuses are in agreement with $M$'s
registration of the `in' count as $n-1$, and no
failure of conjunction introduction has become manifest.

The same
conclusion holds no matter what `in' count $M$ ends up registering.  And if
we think, again, of our
observer, she might well come to believe that not all the marbles are in 
the box by looking at the
pointer of the $M$ apparatus and \emph{not} finding the result `$O=n$'.
Whereas before looking she might have held the belief, for each individual marble, 
that it is in the box, she
could now, if the system quickly evolved to
(\ref{eq:seventeen}), believe that marble 1 is out of the box, each of the others are in the box,
\emph{and} that there are $n-1$ marbles in total in the box. She had no empirical justification for
forming any belief about the `in' count prior to totalling up the marbles.  
But upon totalling them up, she will have to initiate a process that 
amounts to the same thing as setting up the measurement interaction with $M$ and looking at its final
pointer reading.  It will then be a consequence of her instantiating this process that its outcome in fact
agrees  with her
\emph{most current} beliefs
about the individual marbles!  

This conclusion is inescapable even if 
we allow the observer to `form her own opinion' about the marble count 
without bothering to look at the pointer's location on $M$'s dial, \emph{so long as we 
model the observer's opinion forming process within the GRW theory}.  Of course, we do not 
presume to know the contingent details of brain physiology that would be needed 
for a complete GRW model.  However, it suffices to 
assume that whatever brain interactions instantiate `forming opinions 
about marble counts', that process is veridical in the following 
sense.  If, prior to making a judgement about the marble count, the observer 
were in the state
\begin{equation}
\ket{\phi}_{\mbox{out}_{0}}\ket{\mbox{`count?'}}=
     \inket_{1}\ket{\mbox{`in'}}_{M_{1}}\inket_{2}\ket{\mbox{`in'}}_{M_{2}}\cdots 
     \inket_{n}\ket{\mbox{`in'}}_{M_{n}}\ket{\mbox{`count?'}}
     \end{equation} 
---with the different $M_{i}$ states correlated to seperate memory stores in her 
brain, and $\ket{\mbox{`count?'}}$ denoting the initial state of that part of her brain 
that stores arithmetical judgements---then, afterwards, she should be in the state
\begin{equation} 
\ket{\phi}_{\mbox{out}_{0}}\ket{\mbox{`The count is $n$'}};
\end{equation}
 and, 
\emph{mutatis mutandis}, for
\begin{equation}  
\ket{\phi}_{\mbox{out}_{1}}\ket{\mbox{`The count is $n-1$'}}, \ldots,
 \ket{\phi}_{\mbox{out}_{n}}\ket{\mbox{`The count is 0'}}.
 \end{equation}  
Thus, if we make the minimal assumption that our observer would be
 competent to form opinions 
about marble counts when each individual marble is \emph{completely} localized 
(i.e., localized according to the standard eigenstate-eigenvalue link) either 
inside 
or outside of the box, it follows that her brain will instantiate 
exactly the same interaction with the marbles that we have supposed is 
instantiated by $M$, and, hence, that exactly the same conclusions 
that we drew above 
apply.  To put it the other way around: the only way to arrange things 
so that our observer \emph{could} falsify 
the enumeration principle would be to suppose that she was never a 
competent enumerator to begin with!\footnote{To further underscore the point that 
this conclusion is robust 
under differing assumptions about brain physiology, note that our 
considerations remain valid even 
when we suppose that our observer is like Albert's ([1992], Figure 7.15) 
science-fictional character
`John-2'---capable of registering information about the outside world 
in the state of a \emph{single} brain particle.  For, as 
observed in the previous footnote, it is only the sum total of all  
the particles in $M$, the $M_{i}$'s, and the marbles (or particles) 
being counted that needs to be macroscopic.}    

     We have shown that, if one first measures the marbles individually, and then enumerates
the collection as a whole, no failure of conjunction introduction can become manifest; the process
of counting marbles cannot break down. Alternatively, one can consider what happens if one first
measures the system as a whole, and then the marbles individually. After $O$ is measured, the
marbles/$M$-apparatus system
ends up in a state like that of $\ket{\psi}_{\mbox{count}}$, except without the states
of the apparatuses $M_{1}$ through $M_{n}$. As before, since this is now an entangled state,
the system will effectively collapse onto a state such as
\begin{equation} \label{eq:eighteen}
     \outket_{1}\inket_{2}\cdots\inket_{n}\ket{\mbox{`}O=n-1\mbox{'}}_{M}.      
\end{equation}
When the locations of each marble are now measured individually, the system will in all likelihood
end up effectively in state
(\ref{eq:seventeen}), and thus no failure of
conjunction introduction becomes manifest. Since the collapse to (\ref{eq:seventeen}) is only
effective, it is also possible, albeit highly unlikely, for the system 
to further evolve from being effectively
(\ref{eq:seventeen}) to being effectively some other state, like
\begin{equation}
     (\inket_{1}\ket{\mbox{`in'}}_{M_{1}}\cdots
     \inket_{n}\ket{\mbox{`in'}}_{M_{n}})\ket{\mbox{`}O=n\mbox{'}}_{M}     
\end{equation}
so that marble 1 jumps \emph{back} in the box.  But, even in such an 
unlikely scenario, the various apparatuses
will \emph{still} be in agreement at the end of the day; no failure of
conjunction introduction is manifest.  And, again, one could run 
through the same kind of treatment within the brain of an observer. 
     
     By considering cases like this, the general strategy of our argument becomes apparent. To
manifest a failure of conjunction introduction, one has to get an 
(animate or inanimate) apparatus 
which measures the
system as a whole appropriately correlated with the system, and one has 
to get (animate or inanimate) apparatuses which
measure the locations of each marble appropriately correlated with each marble. Once all that is
done, the requisite entanglement between the marbles (or particles) 
will be established and the dynamics of the GRW theory will guarantee that the system will either be in, or almost
instantaneously evolve to,
a state where the various apparatuses are in agreement and no failure of arithmetic is ever
manifest.  The strength of this response to Lewis's counting anomaly 
is that it applies no matter how large we suppose the universe to be, and 
it applies just as well to counting particles as it does to counting 
marbles.

\section{Is Suppressing the Manifestation of Anomalies Enough?}        
We have seen that the GRW theory, together with the fuzzy link, entails that conjunction
introduction can fail for multi-particle systems.  We also noted that, even for a single particle,
there is the anomaly that property intersection can fail.  Moreover, quantum systems can
instantaneously jump between disjoint
regions of space, though for a macrosystem this will virtually never happen.

To this
list, we must also add that full blown action-at-a-distance can be instantiated at the microlevel.
Consider two non-interacting \emph{particles} (not marbles), $L$
and $R$, each of which can either be in
or out of a box, but their boxes are widely separated in space, on the left and right.
Suppose that at some time $t$ their joint state happens to be: \begin{equation} \label{eq:aaad}
a\inket_{L}\inket_{R}+b\outket_{L}\outket_{R},
\end{equation}
with $|b|^{2}\leq p$.  Consider a sufficiently small time interval $T$ around $t$ over which the
free Schr\"{o}dinger evolution of the particles does not invalidate the inequality $|b|^{2}\leq p$. 
We can also suppose that during $T$ the state (\ref{eq:aaad})
does not GRW collapse, because the probability for hits is negligibly low with only two particles
in the system. Then, applying
the fuzzy link to (\ref{eq:aaad}), both particles are determinately in their boxes throughout the
interval $T$.  However, suppose that during $T$ the left-hand
particle were subjected to a measurement of whether it is in or out of its box, producing the state:
\begin{equation} \label{eq:jump}
a\inket_{L}\ket{\mbox{`in'}}_{M_{L}}\inket_{R}+
b\outket_{L}\ket{\mbox{`out'}}_{M_{L}}\outket_{R}.
\end{equation}
Since $M_{L}$ is macroscopic, the probability of GRW collapse during $T$ is now extremely high.  Of course, it is most likely
that an effective collapse to the first term of (\ref{eq:jump}) would occur.  
But it is certainly \emph{not
impossible} that the effective collapse would be to the second term, in which case the particle on
the right, according to the fuzzy link, would have to switch from being determinately in its box to
being determinately out.  Notice that such a switch would have to have 
been brought about
through action-at-a-distance, since in the absence of a measurement interaction on the distant left-hand
particle (the `action'), the right-hand particle (`at-a-distance') would
have remained in its box during $T$.  (We are, of course, well aware 
that even when $|b|^{2}\not\leq p$ there could be a jump in the state of the right-hand
particle; but that jump would not be from one determinate state of affairs 
to another as interpreted via the fuzzy link.)

Is the fact that the GRW theory contains within it `mechanisms' that suppress
 the manifestation of fuzzy link anomalies
sufficient reason to continue to take the theory seriously?  A sceptic might incline towards
the view set forth in Reichenbach's ([1948]) `The Principle of Anomaly in Quantum Mechanics' 
that one should always impose on any interpretation of quantum
theory the requirement that there be no action-at-a-distance behind quantum phenomena.
More generally, Reichenbach appears to argue against attributing an object 
\emph{any} kind of behaviour that is radically different from the way 
the object manifests 
itself to us:
\begin{quote}
Speaking of unobserved objects is meaningful only if such objects are related to observed ones. 
If we say that a tree exists while we do not look at it, or while nobody looks at it, we interpolate
an unobserved object between observables; and we select the interpolated object in such a way
that it allows us to carry through the principle of causality.  For instance, we observe that a tree
casts a shadow; when we see a tree shadow without looking at the tree, we say that the tree is still
in its place and thus satisfy the principle of causality.  More precisely speaking, we select an
interpolation which makes the causal laws of unobserved objects identical with those of observed
ones.  This qualification is necessary because otherwise we could interpolate different objects and
construct for them peculiar causal laws; for example, we could assume that the unobserved tree
splits into two trees, which however cast only one shadow.  It is the \emph{postulate of identical
causality} for observed and unobserved objects which makes statements about unobserved objects
definite\ldots The postulate itself is neither true nor false, but a rule which we use to simplify our
language.  ([1948], p. 341)
\end{quote}
One can ignore the conventionalist overtones of this
passage and still agree that occurrences behind the phenomena should be described, as far as is
possible, in a way that is continuous with the manifest world.

On the other hand, in the course
of a discussion of the confusion between the
instrumental and objective interpretation of wavefunctions, Reichenbach opines: `The confusion
of interpretations is one of the weak spots of the customary discussion of quantum-mechanical
issues; it has blinded the eyes of some physicists to the extent that they do not see the causal
anomalies \emph{unavoidable for every interpretation} [italics 
ours]' ([1948], p. 345).  
Indeed, from what we have said (at the end of Section 3) about no-collapse 
interpretations, it
would not be hard to make a case that \emph{no} interpretation of quantum theory can be
entirely anomaly-free.  The more important issue, it seems to us, is the \emph{status} of an
alleged anomaly, i.e., how \emph{it} should be interpreted.

The GRW theory, considered as a 
theory about the evolution of wavefunctions, is perfectly
consistent with classical logic and arithmetic.
  It is only once we relate wavefunctions to our ordinary language via the fuzzy link
that all the anomalies we have discussed can crop up.  This suggests
that one should sharply distinguish the fundamental ontology of the
GRW theory, viz. wavefunctions evolving and
collapsing in configuration space, from the implications the fuzzy link has for how we are licensed
to \emph{talk} about a world governed by the GRW theory.  Fuzzy link semantics, on this view,
does not add anything of ontological import to the  GRW theory, but simply provides
a way of mapping our `particle' language onto
a theory whose fundamental language concerns wavefunctions.  The fuzzy link, for
some particular value of $p$, would
then have something of the status of a postulate that (to echo Reichenbach 
above) `is neither true nor
false, but a
rule which we use to simplify our language'.

Certainly the argument for this construal of the fuzzy link
(apparently endorsed by Albert and Loewer ([1996], p. 91)) needs to be more fully
developed.
But, supposing it can be, we do not see any reason, in the case of wavefunction collapse
theories, not to answer
`Yes' to the question
in the title of this section.

\begin{center}
 \textbf{Acknowledgements}
 \end{center}
 We would like to thank Yair Guttman and Peter Lewis for helpful 
 discussions, and especially John Norton for rooting out an error of 
 reasoning that occurred at a critical 
 juncture in an earlier draft of this paper.\vspace{.3in}
 
 \noindent \emph{Departments of Philosophy and History and Philosophy of Science, 
 University of
Pittsburgh, Pittsburgh, PA 15260 (e-mail: 
rclifton+@pitt.edu)}.\vspace{.3in}

\noindent\emph{Department of Philosophy, 1879 Hall,
Princeton University, Princeton, NJ 08544-1006 (e-mail: 
bjmonton@princeton.edu)}.\vspace{.3in}

\begin{center}
 \textbf{References}
 \end{center}

\noindent Albert, D. [1992]:
\emph{Quantum Mechanics and Experience},  Harvard: Harvard University 
Press.\vspace{.1in}

\noindent Albert, D. and Loewer, B. [1996]:
`Tails of Schr\"{o}dinger's Cat',  in R.
Clifton (\emph{ed.}), \emph{Perspectives on Quantum Reality}, Dordrecht: Kluwer, pp. 81-91.\vspace{.1in}

\noindent Bell, J. L.  [1986]:
`A New Approach to Quantum Logic', \emph{The British Journal for
Philosophy of Science}, \textbf{37}, pp. 83-99.\vspace{.1in}

\noindent Bohm, D. [1952]: `A Suggested Interpretation of the Quantum Theory in Terms of
``Hidden Variables'', Parts I and II', \emph{Physical
Review}, \textbf{85}, pp. 166-93.\vspace{.1in}

\noindent Clifton, R.  [1996]:
`The Properties of Modal Interpretations of Quantum Mechanics', 
\emph{The British Journal for
Philosophy of Science}, \textbf{47}, pp. 371-98.\vspace{.1in}

\noindent Ghirardi, G. C. and A. Bassi  [1999]:
`Do Dynamical Reduction Models Imply That Arithmetic Does Not Apply to Ordinary
Macroscopic Objects?',
\emph{The British Journal for Philosophy of Science}, \textbf{50}, pp. 705-20.  \vspace{.1in}

\noindent Ghirardi, G. C., R. Grassi, and P. Pearle  [1990]: `Relativistic Dynamical Reduction
Models: General Framework and Examples',
\emph{Foundations of Physics}, \textbf{20},
pp. 1271-316.\vspace{.1in}

\noindent Ghirardi, G. C., A. Rimini, and T. Weber  [1986]: `Unified Dynamics for Microscopic
and Macroscopic Systems', \emph{Physical Review D}, \textbf{34},
pp. 470-91.\vspace{.1in}

\noindent Hegerfeldt, G. C.  [1995]: `Problems About Causality in 
Fermi's Two-Atom Model and Possible Resolutions', in H. -D. Doebner, 
V. K. Dobrev, and P. Natterman (\emph{eds}), 
\emph{Nonlinear, Deformed and Irreversible Quantum Systems},  
Singapore: World Scientific.\vspace{.1in}

\noindent Kochen, S. and E. P. Specker [1967]: `The Problem of Hidden Variables in Quantum
Mechanics', \emph{Journal of Mathematics and Mechanics}, \textbf{17}, 
pp. 59-87.\vspace{.1in}

\noindent Lewis, P.  [1997]:
`Quantum Mechanics, Orthogonality, and Counting',
\emph{The British Journal for Philosophy of Science}, \textbf{48}, 
pp. 313-28.\vspace{.1in}

\noindent Reichenbach, H.  [1948]:
`The Principle of Anomaly in Quantum Mechanics',
\emph{Dialectica}, \textbf{2},
pp. 337-49.\vspace{.1in}

\noindent Thaller, B. [1992]:
\emph{The Dirac Equation}, Berlin: Springer.\vspace{.1in}

\end{document}